# Thermal Behavior of Astrophysical Amorphous Molecular Ices

Murthy S. Gudipati[1], Benjamin Fleury[2], Robert Wagner[3], Bryana L. Henderson[1], Kathrin Altwegg[4], Martin Rubin[4]

[1]Science Division, Jet Propulsion Laboratory, California Institute of Technology, 4800 Oak Grove Drive, Pasadena, CA 91109, USA. [2]Universite Paris Est Creteil and Universite Paris Cité, CNRS, LISA, F-94010 Créteil, France, [3]Institute of Meteorology and Climate Research, Karlsruhe Institute of Technology, Karlsruhe, Germany, [4] Physikalisches Institut, University of Bern, Sidlerstrasse 5, CH-3012 Bern, Switzerland.

## Abstract

Ice is a major component of astrophysical environment – from interstellar molecular clouds through protoplanetary disks to evolved solar systems. Ice and complex organic matter coexist in these environments as well, and it is thought primordial ice brought the molecules of life to Earth four billion years ago, which could have kickstarted origin of life on Earth. To understand the journey of ice and organics from their origins to becoming a part of evolved planetary systems, it is important to complement high spatial and spectral resolution telescopes such as JWST with laboratory experimental studies that provide deeper insight into the processes that occur in these astrophysical environments. Our laboratory studies are aimed at providing this knowledge. In this article we present simultaneous mass spectrometric and infrared spectroscopic investigation on how molecular ice mixtures behave at different temperatures and how this information is critical to interpret observational data from protoplanetary disks as well as comets. We find that amorphous to crystalline water ice transformation is the most critical phenomenon that differentiates between outgassing of trapped volatiles such as $CO_2$ vs. outgassing of pure molecular ice domains of the same in a mixed molecular ice. Crystalline water ice is found to trap only a small fraction of other volatiles (<5%), indicating ice grain composition in astrophysical and planetary environments must be different depending on whether the ice is in amorphous phase or transformed into crystalline phase, even if the crystalline ice undergoes radiation-induced amorphization subsequently. Crystallization of water ice is a key differentiator for many ices in astronomical environments as well as in our Solar System.

**Introduction**

Astrochemistry is aimed at simulating astrophysical environments to predict chemical processes as well as providing their underlying mechanisms in these environments. As JWST continues to provide outstanding data at better spatial and spectral resolution and sensitivity than the previous telescopic measurements [Ballering et al. 2021; Hinkley et al. 2022; Jin et al. 2022; Muller et al. 2022], the need to obtain laboratory data at higher resolution is now more important than ever before. In addition to higher spectral resolution that enables better separation of individual absorption/emission bands and higher spatial resolution of observations that provides better contrast of the observed objects (whether interstellar molecular clouds or protoplanetary disks or other solar system objects), we need a better understanding of the effect of temperature, pressure, and local radiation in these environments that are often spatially heterogeneous. Small temperature variations could have a significant effect on the ice vs. gas-phase composition in these environments [Baillie et al. 2015; Lecar et al. 2006]. Temperature measurements in astrophysical environments are still averaged through the line of sight and often need to be approximated. For this reason, if laboratory studies could provide better handle on the effect of temperature on the ice/gas partitions, we should be able to use these data to better constrain temperatures of local astrophysical environments. Observations with higher spatial resolution can then be better interpreted using laboratory data.

In protoplanetary environments, ice-lines are drawn at various radial and lateral distances from the protostar [Baillie et al. 2015; Owen 2014; van der Marel et al. 2019]. However, these lines are dictated by not only ice composition and its phase (amorphous or crystalline), but also the exact temperature of the local environment. For this reason, we focused in our study on outgassing of mixed molecular ices and their composition at various temperatures. In protoplanetary disks, though temperature decreases with radial distance, volatile and supervolatile molecular species (with respect to water) would form pure ices or mixed ices trapped in water and other ices based on their local abundances. It is still not clear whether ice particles form in the protoplanetary disks freshly or retain their interstellar ice grain properties as they accrete on to a protoplanetary disk during its formation. Though radial transportation is used in many models, how far the grains and gas are transported and whether transport occurs primarily in one direction or both directions (high to low and low to high temperatures) are still open questions [Aikawa 2005; Bosman et al. 2018], particularly, when gas-phase observations are used as a proxy for ice composition [Walsh et al. 2014]. It is also assumed that the icy grains in the protoplanetary disks become a part of cometary nuclei retaining their chemical, structural, and compositional integrity [Alexander et al. 2017; McClure et al. 2015; Oberg et al. 2015]. Thus, there is an effort to correlate cometary outgassing (reflecting thermal evolution of a comet during its travel around the Sun) with the ice grain composition of protoplanetary disks. In order to gain better constraints on interpreting observational data from icy environments – be they interstellar, protoplanetary, or cometary nuclei - we need a wide range of laboratory data. We conducted temperature programmed desorption (TPD) studies of mixed molecular ices by means of simultaneous mass spectrometry of outgassed molecules and infrared spectroscopy of mixed molecular ices. We also carried out calorimetric study of macroscopic water ice. Our experiments provide some of these critically needed data and understanding of the processes involved, complement existing experimental data as discussed in the following, and provide a deeper understanding of the role of local temperature on the ice physics.

**Experimental:**

All the experiments were conducted at the Ice Spectroscopy Laboratory (ISL) of the Jet Propulsion Laboratory. We used standard Temperature Programmed Desorption (TPD) methods to determine sublimation of mixed molecular ices at various starting mixing ratios. The experimental station (named Himalaya) used for this work is shown in **Figure 1**.

**TPD of Mixed Molecular Ices:** Molecules used here are CO, $O_2$, $CO_2$, and $H_2O$ as a realistic representative of supervolatiles (CO and $O_2$), volatiles ($CO_2$) in water ice. We chose much higher mixing ratio such that the volatiles are in "excess" compared to what has been observed in astrophysical or cometary ices [Mousis et al. 2021; Oberg et al. 2011; Piso et al. 2016; Rubin et al. 2020]. This was deliberately done to ensure that there are pure volatile and supervolatile ice domains in the mixed molecular ices. This method differs slightly from previous methods [Malyk et al. 2007; Martin-Domenech et al. 2014], where either observed mixing ratios were used for mixed molecular ices or layered ices [Acharyya et al. 2007; Bar-Nun et al. 2013; Collings et al. 2004; Collings et al. 2003; Collings et al. 2005; Edridge et al. 2013; Fayolle et al. 2011; Fraser et al. 2004; Fuchs et al. 2006; Greenberg et al. 2017; Muller et al. 2018; Rawlings et al. 2006; Taj and McCoustra 2020], or pure ices were studied through TPD [Isokoski et al. 2014; Mate et al. 2008]. In astrophysical environments it is unlikely that pure supervolatile ice grains exist where no other molecule is present in the ice grains, such as pure CO ice, though pure CO ice domains could exist in an ice grain that contains other volatiles as well. Properties of such mixed-molecular ices or layered ices also depend upon whether these ice grains travelled from cold to warmer regions of the protoplanetary disk or vice versa. It is easy to imagine grains moving from inner protoplanetary disk to the outer, which would lead to less volatile molecules to condense first forming layered ice grains with supervolatiles condensing on the top of the ice grain. The situation is more complex if the movement is from colder to warmer regions as the supervolatiles that are trapped within the less volatile molecular ices only survive in that ice grain. For this reason, we think our molecular mixing ratio provides a more realistic representative of pure ice domains in mixed molecular ice grains and complements data so far available on other ice compositions mentioned earlier.

All the ice mixtures were deposited at 10 K on a sapphire substrate at normal incidence for a duration of 3-6 min at a background pressure of ~1x10$^{-6}$ mbar. The end of the vapor deposition tube (1/4-inch outer diameter stainless steel) was kept ~ 2 inches away from the sapphire optical window that was cooled with a Sumitomo (SHI-4-15, Janis Research) closed-cycle helium cryostat that reaches a minimum of 4.2 K. Calibration studies done on water ice under these conditions show the ice films are of ~250 nm (about 1000 monolayers) in thickness. Initial mixing ratios were controlled based on pressure in the ice chamber determined by an inverted magnetron pressure gauge (Agilent Varian IMG-300). We did not use mass spectrometer to adjust the mixing ratios because of fragmentation of $CO_2$ to CO that interferes with exact ratios. Final mixing ratios are determined from the FTIR spectral integrated intensities normalized to the respective bandstrengths (see results section) of the individual molecules.

The temperature ramp was held at 0.5 K/min. Our Lakeshore (Model 340) temperature controller allows for a ramp rate as slow as 0.1 K/min, but the long duration experiments introduce artefacts such as background deposition that may lead to erroneous results. We experimented with faster ramp rates (1 K/min) and found that the rate did not show significant effect on the TPD-mass spectrometry (TPD-MS) of these ices. Most of the experiments were done with 0.5 K/min heating rate to ensure better equilibration of ice at any given temperature and better applicability of TPD data to typical astrophysical environments.

Desorption of molecular volatiles during TPD was quantified through quadrupole ion mass spectrometry (MS) using Stanford Research Systems (SRS RGA-300) quadrupole mass spectrometer and infrared (IR) spectroscopy using Thermo-Nicolet 6700 FTIR spectrometer equipped with KBr beam splitter and external liquid-nitrogen cooled MCT-A detector. IR spectroscopy provides independent and complementary data on the desorption process in terms of molecules remaining in the ice phase, whereas mass spectrometry provides composition of desorbed molecules in vacuum. We found that this combination of techniques was particularly useful in cases where species are less-efficiently pumped by the turbomolecular pump (i.e., CO molecules) and thus can be overrepresented in the mass spectra, or for those that are unable to be quantified by the infrared spectra alone (i.e., $N_2$ and $O_2$ molecules). With IR spectroscopy, we were also able to track a slight increase in the IR absorption of the $H_2O$ ice band as the temperature is ramped due to sublimation from other parts of the cryostat and recondensation on the sapphire window.

Ionization of gases in the mass spectrometer was done at 70 eV. Electron-impact ionization cross-sections at 70 eV for these molecules are taken from NIST data ($O_2$ = 2.441 $Å^2$; CO = 2.516 $Å^2$; $CO_2$ = 3.521 $Å^2$; $H_2O$ = 2.275 $Å^2$). Integrated mass spectral data were then normalized with respect to the $CO_2$ ionization cross-section to obtain molecular mixing ratios of deposited gas mixture (which in turn represents closely the ice composition determined through infrared spectroscopy). The scaling factors used are: $O_2$ (1.44); CO (1.4); $CO_2$ (1); $H_2O$ (1.55). We note here that CO mass peak at 28 m/z has contribution from $N_2$ (which can be estimated to be about 8 times the partial pressure of N detected at 14 m/z) as well as ionization fragmentation of $CO_2$ to CO, which is about 10% of $CO_2$ partial pressure. These and slow pumping of CO by turbomolecular pumps cause difficulties to quantitatively interpret CO mixing ratio in the outgassing under our experimental conditions.

Using the integrated IR spectral absorbances, the relative molecular abundance was calculated using bandstrengths from the literature [Gerakines and Hudson 2015; Gerakines et al. 1995; Giuliano et al. 2014; Hudgins et al. 1993] as follows: CO (1.1E-17 cm/mol); $CO_2$ (7.7E-17 cm/mol); $H_2O$ (1.6E-16 cm/mol). Please note that $O_2$ is IR inactive, and we rely upon mass spectrometry for $O_2$ mixing ratios. Further, we conducted calorimetry using the same conventional closed-cycle cryostat that is also used for the above mentioned TPD studies.

**Calorimetry of Pure $H_2O$ Ice:** For the calorimetry experiments, we deposited macroscopic amounts of pure water at 10 K to form amorphous ice. Our goal was to capture the amorphous to crystalline phase transition of pure $H_2O$ ice and determine if there were differences in temperature range or any dependence on ice thickness on the crystallization of amorphous ice during TPD. For these experiments, we made an internal calibration between two temperature measuring silicon diodes, first diode (SD1) is mounted at the end of the heating tape where the sample holder is connected to the closed-cycle cryostat and the second silicon diode (SD2) is mounted at the bottom of the sample holder (**see Figure 2**). Amorphous ice was directly deposited on the copper sample holder where the SD2 is located for different durations. The water vapor inlet tube was brought about 5 mm closer to the SD2 location of the sample holder to avoid ice formation on the entire sample holder, but to limit it at the SD2 location. First, we ran a pre-programmed 2 K/min reference ramp prior to ice deposition using the Lakeshore temperature controller's temperature ramping with PID (proportional, integral, derivative) control. To obtain a smooth heating ramp devoid of the oscillations inherent in the PID control, we fitted the recorded time-dependent heater wattage values with a three-exponential fit and used these values to prescribe the heating ramp for the ice experiments.  If there were no change in the heat absorbed or released by the ice, at each pre-programmed ramping stage, the temperatures of both the diodes SD1 and SD2 would have read the same as without the sample. However, if there were to be heat released due to the exothermic amorphous to crystalline phase transition, the SD2 would read a higher temperature than SD1 and the ratio SD2/SD1 would increase. Similarly, for an endothermic phase transition, this ratio would decrease. Indeed, subtle changes were measured corresponding to exothermic ice crystallization around 140 K and endothermic ice sublimation around 190 K. The ice samples deposited were estimated to be a few milligrams and the heating ramp was conducted at the maximum power setting on the heater controller at 100 W, starting at 35% of the output at 10 K and ending at ~90% at 200 K. Despite the high heater wattage needed to work against the cooling power, we were able to measure extremely small exothermic and endothermic peaks because we have used the two temperature sensor diodes located apart from each other that would otherwise have exactly the same conditions except for instantaneous minuscule differences in temperature. This is almost like weighing an elephant with and without a sparrow on it! But it indeed worked quantitatively as will be presented in the results section.

**Results**

  **TPD of Mixed Molecular Ices:** We studied several mixed molecular ices, including two component $CO:CO_2$, three component $CO:O_2:CO_2$, and four component $CO:O_2:CO_2:H_2O$ ices. We also varied the mixing ratios, except the CO to $O_2$ ratio was kept constant close to 1. In the TPD-MS, we followed gas-phase N (proxy of $N_2$ that comes at the same m/z as CO at 28; N:N2 peak ratio from NIST Webbook is ~12:100), CO, $O_2$, $CO_2$, $CO_3$ (that could be an impurity in CO gas cylinders through reaction of CO with $O_2$), and $H_2O$. Some of these desorption mass spectral plots are shown in **Figures 3 – 7** along with their counterpart from integration of the FTIR spectra. In **Figure 8,** integrated mass spectral data, of outgassed volatiles trapped in amorphous water ice (>90 K) that are released during amorphous to crystalline ice transition ~140 K (black bars) and trapped in crystalline water ice (>145 K) that are released along with subliming water (orange bars), are shown as bar graphs. Full range mass spectra and infrared spectra for four component mixed molecular ice are shown in **Supplementary Figures S10 and S11**. Our TPD-MS profiles are qualitatively in agreement with earlier work, especially pure and mixed-molecular ices containing $H_2O$, CO, $CO_2$, $CH_3OH$, and $NH_3$ [Martin-Domenech et al. 2014] and $H_2O$, CO, $CO_2$, $CH_4$ [Kouchi and Yamamoto 1995]. In all the experiments we conducted, most of the CO either exists as a pure CO ice domain or CO is outgassed prior to the start of $CO_2$ outgassing, even if CO is trapped in $CO_2$ ice, except with $CO:O_2:CO_2:H_2O$ mixed molecular ices. During the formation of water-dominant amorphous ice, all the other volatiles are trapped to some extent within the amorphous ice. We also observed some entrapment of $CO_2$ in pure CO or $O_2$ ice domains evident from $CO_2$ outgassing along with CO and $O_2$, both of which have similar sublimation temperatures. $N_2$ sublimation starts after CO sublimation and along with $O_2$ sublimation, as inferred from tracing N in the TPD-MS profiles. Also, minor entrapment of CO and $O_2$ was observed in $CO_2$ ice domains as reflected in the $CO_2$ sublimation region of the TPD-MS profiles. Above 100 K, when all the excess pure ices of CO, $O_2$, and $CO_2$ are sublimed, the remaining ice on the substrate must be amorphous water ice and any residual volatiles/supervolatiles trapped in it. Indeed at ~130 K, we start observing another desorption event that maximizes at 145 K and completes at 150 K. During this period, a significant amount of all three volatiles CO, $O_2$, and $CO_2$ are released simultaneously into the gas-phase, along with a small amount of $H_2O$ sublimation. We see consistently that more $CO_2$ is trapped in amorphous water ice than CO or $O_2$. From the integrated TPD-MS data we find that the sum of all trapped species in amorphous water ice is about 30%. Our data indicates that indeed amorphous water ice can trap significant amounts of other volatile species even at temperatures as high as 130 K. Interestingly, as soon as the temperature reaches 150 K all the trapped volatiles are desorbed and the amorphous to crystalline ice transformation is then complete as well. Above 150 K, water ice starts sublimation alone. At this point only an extremely small portion of other volatiles are trapped in the crystalline water ice, as evident from the desorption maxima around 160 K, where only about 2-3% of CO is seen to desorb along with $H_2O$. The rest of the molecules have even lower mixing ratios. We can use the TPD-MS and IR data of our experiments to backtrack ice composition at different temperatures.

Infrared spectra have also been simultaneously obtained and the discrepancies between IR and mass spectral data reveal values obtained for CO from mass spectral data alone cannot be quantitatively used due to the longer residence time of CO in the vacuum chamber. For example, total integrated TPD-MS partial pressure ratios of CO:$CO_2$ is ~3:1, whereas IR spectral data indicate a CO:$CO_2$ ratio of 1:2.8 (**Figure 3**). It is well known that conventional turbomolecular pumps even at $10^{-9}$ mbar vacuum are less efficient to pump CO. As a result integrated mass spectra overestimate CO outgassing. This is further complicated by continuous CO release from the ice into the vacuum that is clearly seen from the IR spectra of the ices (**Figures 3-7**). In the astrochemistry community, the majority of the cryogenic systems are equipped with these turbomolecular pumps, at ~$10^{-9}$ mbar working pressures. As discussed below, under those conditions caution must be exercised in the interpretation of TPD-MS data alone.

**Calorimetry of Pure $H_2O$ Ice:** In order to quantify the exothermicity of the amorphous to crystalline ice transition, we have relied on an internal standard, namely the heat of sublimation of crystalline ice. In our first experiment, where ice was deposited at 2-3*$10^{-5}$ mbar for 900 min at a rate of 1*$10^{-5}$ g/min (black line curve in **Figure 9**), the endothermic sublimation as well as the exothermic crystallization are well behaved. Subsequently, we increased the time of deposition as shown in **Figure 9** and both the exotherm and endotherm peaks increased, with the sublimation endotherm out of bound with thicker ice sample. At this point, ice sublimation is so fast that the vacuum in the system raised significantly (to ~1E-3 mbar) and the data was no longer reliable. We excluded such data from calculating sublimation enthalpy. From the literature, we took the sublimation endotherm to be 2830 kJ/kg [Feistel and Wagner 2007] and multiplied it by the ratio of the integrated crystallization exotherm to the integrated sublimation endotherm. By this procedure, we obtained the heat of crystallization exotherm for amorphous ice to be 109, 109, 116, and 138 kJ/kg for the four different experiments we conducted at different ice deposition durations ranging from 900 min to 5160 min, with an arithmetic mean and standard deviation of 118 ±12 kJ/kg. This value is in very good agreement with the literature reports of ~100 kJ/kg [Chonde et al. 2006; Handa et al. 1986; Kouchi and Sirono 2001; Smith et al. 2011]. Our method is simple and easy to implement for a wide range of equipment and ices.

**Discussion:**

There are a few major take aways from our experiments presented here. The first one is that super volatiles CO, $O_2$, and $N_2$ (proxy N) can only exist in significant mixing ratios in interstellar, protoplanetary, and cometary ices that are extremely cold – at or less than 30 K. The second take away is that the exothermicity of amorphous to crystalline pure water ice transition is confirmed again. At the same time, we also found that irrespective of the ice thickness – whether a few hundred nanometers thin ice films or few hundred micrometers to millimeters thick ice grains – the amorphous to crystalline ice transition occurs within a narrow temperature range of 130 K to 150 K, with a maximum at ~140 K. Another experimental take away is that quantification of CO in the gas-phase under vacuum should be carefully conducted and should be verified by complementary experiments. Finally, conventional cryogenic systems with no additional instrument modification can be used to carryout calorimetry of ices and we plan to extend our calorimetric studies to mixed-molecular ices in the future. In the following we discuss more in detail temperature ranges and their effect on the ice/gas composition.

**Below 30 K:** Most interstellar ices, protoplanetary ices at greater radial distances, and Oort Cloud comet interiors are expected to be at or below 30 K. Under these conditions, based on our data presented here, we can expect no outgassing of any supervolatiles such as CO, $O_2$, $N_2$ ($CH_4$, Ar etc., by inference) either as pure ices or trapped in mixed molecular ices. So, supervolatile-rich ices should be present at temperatures below 30 K. We can approximately refer to this the CO ice line.

**30 – 75 K:** This is the temperature range one expects in protoplanetary disks at closer distances than the CO ice line mentioned above. These temperatures are also present as equilibrium temperatures of solar system icy bodies such as TNOs (Trans Neptune Objects) and Kuiper Belt Objects (KBOs), and Centaurs [Guilbert-Lepoutre et al. 2015; Guilbert-Lepoutre et al. 2023]. In these environments, the ice grains are expected to be depleted of significant amounts of supervolatiles CO, $O_2$, and $N_2$ ($CH_4$, Ar etc., by inference), but still contain significant amounts of $CO_2$. For this reason, Jupiter family short-period comets that originate from KBOs and Centaurs should be significantly depleted of CO and other supervolatiles, but still retain high amounts of $CO_2$. In the protoplanetary disks, the ice grains accreting from the interstellar environment and moving radially into the inner disk would have significant amounts of $CO_2$, but otherwise be depleted in supervolatiles. From the TPD-MS and TPD-IR data presented here, we find that though significant amount of CO is outgassed early (below 50K), CO continues to leave ice-phase gradually until $CO_2$ starts sublimation. This indicates that some of the CO molecules, either trapped in $CO_2$ or existing as a CO ice domain covered by $CO_2$ ice, outgas at higher temperature. We should be cautious to over interpret this data, as this outgassing is expected to be both temperature and time dependent.

**75 – 120 K:** This temperature range leads to the loss of any excess pure $CO_2$ ice and any volatiles left behind are now completely trapped within amorphous water ice cages in the 4-component ices. In three and two component ices, where $CO_2$ is the dominant species, we observe very small amounts (<3%) of supervolatiles (including $N_2$, $O_2$, and CO) to outgas along with $CO_2$ (**Figures 3-5**), indicating $CO_2$ ice has crystallized expelling other impurities well before its sublimation onset [Escribano et al. 2013]. Centaurs, the precursors of Jupiter-family comets, could have their equilibrium temperatures in this range [Drahus et al. 2017; Jewitt 2009]

**120 – 150 K:** This is the most critical temperature in astrophysical environments, and it is a point of no return, when amorphous water ices irreversibly undergo phase transition into crystalline water ice, during which process all the volatiles trapped within the amorphous ice are released. Our experiments show that about 17% $CO_2$, ~9% CO, and ~4% $O_2$ are trapped in - warm amorphous ices beyond 90 K – and these are released during ice crystallization between 130 K and 150 K. These values are close to the modeling work done recently to reproduce comet 67P/Churyumov-Gerasimenko nucleus properties [Davidsson et al. 2022]. It is expected that crystallization of amorphous water ice is slower at lower temperatures and faster as the temperatures approach 150 K [Mastrapa et al. 2013]. This temperature region in protoplanetary disks should be narrow but clearly marked by crystalline water-ice features in the IR spectral observations. JWST should be able to detect these regions and constrain the temperatures. Comets should also show unusual $CO_2$-rich outgassing and $H_2O$ inactivity in these temperature ranges. Indeed, such dichotomy is observed in the outgassing of many comets, including 67P/CG [Fink et al. 2016; Herny et al. 2021; Hoang et al. 2019; Tubiana et al. 2019], whereby subsolar regions with higher temperatures show $H_2O$ outgassing (crystalline water ice sublimation – see below), whereas other regions with lower temperatures show enhanced $CO_2$ outgassing. It is very likely that amorphous water ice loaded with $CO_2$ and other volatiles, at depths where thermal equilibrium reaches ~130 K, eject these volatiles while undergoing crystallization. As a result, a significant amount of $CO_2$, CO, and other volatiles are observed from the regions where the surface temperatures are lower. Once crystallized, water ice from the same region, when temperatures are higher due to direct exposure to Sun, results in predominantly $H_2O$ sublimation (and depleted in $CO_2$). Based on this laboratory observation alone, we can confidently conclude that cometary nuclei are predominantly made of amorphous water ice, at least deep inside the nucleus. We also note that presence of silicate dust and refractory organic mantle on short-period comets could further effect outgassing from comets [Herny et al. 2021; Hoang et al. 2019; Hoang et al. 2020].

**150 – 190 K:** Above 150 K crystalline water ice starts to sublimate rapidly. Based on the thickness of the crystalline water ice film, sublimation could be complete at 170 K or above. The sublimation temperature also depends on the ambient pressure of the ice, with higher pressures shifting sublimation to higher temperatures. Such a scenario does not exist in low-pressure protoplanetary environments but could be present in the near-subsurface ices of the outer Solar System icy bodies and cometary near-subsurface ices, as a comet gets closer to its perihelion. The location of crystalline water ice sublimation in the protoplanetary disks is also designated as $H_2O$ ice line that is closer to 150 K. However, based on the above-mentioned variables, it is possible that water ice could persist at higher temperatures.

If an accreted interstellar amorphous ice grain makes its journey through each of the ice lines in a protoplanetary disk, shedding away volatiles, the remaining material on these grains would be highly refractory (silicates and complex organics). If these grains or silicate dust from the inner protoplanetary disk were subsequently to move radially outwards, then the ice condensation on these grains will be layered (onion-like), with the non-volatiles such as $H_2O$ condensing first, followed by other volatiles such as $CO_2$, and finally the supervolatiles such as CO. Thus, the composition of an ice grain – layered vs. intimately mixed – depends on whether it was formed within the protoplanetary disk or retained from interstellar grains.

The structural composition of these grains (layered vs. intimately mixed) plays an important role in the subsequent chemistry of these ices. Because highly energetic particles can deposit energy in ionization tracks as they penetrate through matter, reactions tend to occur with nearby species. As a result, any radiation-induced chemistry of protoplanetary- disk -formed ice grains should be different from interstellar ice grain chemistry.

Cometary nuclei have been puzzling researchers for decades. Recent Rosetta mission provided more insight into the outgassing properties of comet 67P/CG [Herny et al. 2021]. High porosity of cometary nuclei and large thermal gradients make predictions of cometary nucleus composition based on outgassing difficult. However, based on our data and earlier experimental work, we can provide some insight into cometary nuclei: (a) If the entire nucleus of a comet reaches its equilibrium temperature above 70 K, it would be unlikely that these comets contain regions of pure molecular ices of super volatiles such as CO, $N_2$, $CH_4$, etc. We observe continuous outgassing of CO from the mixed molecular ice above 30 K until $CO_2$ sublimation starts at 70 K. More laboratory work is needed to understand long-term effect of temperature between 30 K and 70 K on the outgassing of these supervolatiles to determine if any of their pure ice domains would survive these temperatures. $CO_2$ ice domains are more likely to survive in the interiors of a comet nucleus with equilibrium temperature at or below 70 K, followed by trapped volatiles in amorphous $H_2O$ ice below ~120 K. At closer heliocentric distances, where surface temperatures raise beyond $H_2O$ ice sublimation temperatures below the dust mantle, outgassing properties of a comet would be most complex due to thermal wave propagating deeper that results in not only sublimation of crystalline water ice, but also outgassing from amorphous to crystalline phase transition at deeper regions releasing significant amount of trapped volatiles (also known as molecular volcanos on a comet)[Collings et al. 2004; Smith et al. 1997a; Smith et al. 1997b]. However, at even greater depths $CO_2$ ice domains could start sublimation reaching the surface. However, outgassing dichotomy of comets indicates that $CO_2$ is predominantly from relatively lower temperature regions on surface compared to $H_2O$, indicating $CO_2$ outgassing could have contribution from all these processes. At larger heliocentric distances, however, $H_2O$ amorphous-to-crystalline transition and sublimation are significantly suppressed, and it is expected that $CO_2$ ice domains dominate the outgassing in these regions. Rosetta mission also found strong correlation between $H_2O$ and $O_2$ outgassing, up to 5% mixing ratio of $O_2$ [Bieler et al. 2015; Heritier et al. 2018]. In our experiments, we see much smaller fraction of $O_2$ outgassing along with $H_2O$ (see Figure 8). One possibility is that cometary nuclei and their precursors may have been irradiated by cosmic rays over the past 4.6 billion years, resulting radiation-induced trapping of $O_2$ in $H_2O$ ice as it has been seen on the surface of Galilean moons Europa, Ganymede, and Callisto [Spencer et al. 1995] and experimentally demonstrated during ion irradiation of ice [Teolis et al. 2009]. More laboratory studies are needed to understand complex thermal properties of cometary nuclei and their outgassing properties during their journey around the Sun.

In the outer Solar System, on icy satellites such as Europa and other Jovian and Saturnian moons, surface temperatures can vary between 150 K and 60 K depending on latitude and sub-solar regions [Ashkenazy 2019; Berdis et al. 2020; Oza et al. 2019; Rathbun et al. 2010]. Sub-solar equatorial regions sustain the highest possible temperatures, whereas polar regions remain cooler. Similarly, day and night temperatures also vary significantly. Under these conditions, water ice sublimation occurs from the crystalline ice phase into vapor, forming a tenuous atmosphere. When the water vapor condenses on the surface, amorphous water ice could form in cold higher latitude regions or at lower latitudes on the night side. At temperatures below 70 K, this refrosting process could also lead to trapping of $CO_2$ if it is present in the atmosphere.

**Conclusions**

Temperature plays an important role in the compositional properties of astrophysical ices. Laboratory data under various conditions will provide better constraints on the interpretation of observational data, whether it is protoplanetary disks, comets, or other Solar System icy bodies. $H_2O$ can co-condense with significant amounts of other volatiles and supervolatiles can at very low temperatures (<30 K), whereas at warmer – above the $CO_2$ sublimation – temperature (>90 K), amorphous water ice has only a limited ability to trap volatiles, of which $CO_2$ has the highest mixing ratio. Crystallization of amorphous ice expels all the trapped volatiles around 130-150 K. This temperature region is the same, whether the ice grains are thin (a few hundred nanometers) or thick (a few hundred microns), based on our differential calorimetry experiments. Such a process could lead to outgassing of significant amounts of $CO_2$, whereas $H_2O$ is still in the solid form. This important process likely explains the outgassing dichotomy observed in comets where $CO_2$ outgassing occurs from a different region of the comet than $H_2O$ outgassing, which is predominantly from the subsolar regions. We also found that crystalline water ice traps only a small fraction of other volatiles (<5%), indicating ice grain composition in astrophysical and planetary environments must be different depending on whether the ice is in amorphous phase or transformed into crystalline phase. Once irreversibly transformed into crystalline phase, water ice is expected not to trap anymore volatiles, even if the crystalline ice undergoes radiation-induced amorphization subsequently. As for the protoplanetary disks, any radial mixing over long distances from the protostar could lead to entirely different ice composition, namely layered ices, instead of the mixed molecular ices formed in the interstellar molecular clouds. The morphology of these layered ices could also lead to different radiation chemistry compared to mixed molecular ices. JWST observations should provide better insight and constraints on the formation and mobility of ice grains in protoplanetary environments.

## Acknowledgments

This research work was carried out at the Jet Propulsion Laboratory, California Institute of Technology, under a contract with the National Aeronautics and Space Administration. This research was enabled through funding to MSG from NASA SSW and DDAP Programs.

## References:

Acharyya, K., et al. (2007). "Desorption of CO and o-2 interstellar ice analogs." Astronomy & Astrophysics **466**(3): 1005-U1169.
Aikawa, Y. (2005). Chemistry in protoplanetary disks. Highlights of astronomy, vol 13. O. Engvold. San Francisco, Astronomical Soc Pacific. **13:** 515-517.
Alexander, C. M. O., et al. (2017). "Measuring the level of interstellar inheritance in the solar protoplanetary disk." Meteoritics & Planetary Science **52**(9): 1797-1821.
Ashkenazy, Y. (2019). "The surface temperature of europa." Heliyon **5**(6).
Baillie, K., et al. (2015). "Time evolution of snow regions and planet traps in an evolving protoplanetary disk." Astronomy & Astrophysics **577**.

Ballering, N. P., et al. (2021). "Simulating observations of ices in protoplanetary disks." Astrophysical Journal **920**(2).

Bar-Nun, A., et al. (2013). Gas trapping in ice and its release upon warming. The science of solar system ices. M. S. Gudipati and J. Castillo-Rogez, Springer New York. **356:** 487-499.

Berdis, J. R., et al. (2020). "Europa's surface water ice crystallinity: Discrepancy between observations and thermophysical and particle flux modeling." Icarus **341**.

Bieler, A., et al. (2015). "Abundant molecular oxygen in the coma of comet 67p/churyumov-gerasimenko." Nature **526**(7575): 678-681.

Bosman, A. D., et al. (2018). "Efficiency of radial transport of ices in protoplanetary disks probed with infrared observations: The case of CO2." Astronomy & Astrophysics **611**.

Chonde, M., et al. (2006). "Glass transition in pure and doped amorphous solid water: An ultrafast microcalorimetry study." Journal of Chemical Physics **125**(9).

Collings, M. P., et al. (2004). "A laboratory survey of the thermal desorption of astrophysically relevant molecules." Monthly Notices of the Royal Astronomical Society **354**(4): 1133-1140.

Collings, M. P., et al. (2003). "Carbon monoxide entrapment in interstellar ice analogs." Astrophysical Journal **583**(2): 1058-1062.

Collings, M. P., et al. (2005). "Sub-monolayer coverages of CO on water ice." Chemical Physics Letters **415**(1-3): 40-45.

Davidsson, B. J. R., et al. (2022). "Modelling the water and carbon dioxide production rates of comet 67p/churyumov-gerasimenko." Monthly Notices of the Royal Astronomical Society **509**(2): 3065-3085.

Drahus, M., et al. (2017). "New limits to CO outgassing in centaurs." Monthly Notices of the Royal Astronomical Society **468**(3): 2897-2909.

Edridge, J. L., et al. (2013). "Surface science investigations of the role of CO2 in astrophysical ices." Philosophical Transactions of the Royal Society a-Mathematical Physical and Engineering Sciences **371**(1994).

Escribano, R. M., et al. (2013). "Crystallization of CO2 ice and the absence of amorphous CO2 ice in space." Proceedings of the National Academy of Sciences of the United States of America **110**(32): 12899-12904.

Fayolle, E. C., et al. (2011). "Laboratory H2O:CO2 ice desorption data: Entrapment dependencies and its parameterization with an extended three-phase model." Astronomy & Astrophysics **529**.

Feistel, R. and Wagner, W. (2007). "Sublimation pressure and sublimation enthalpy of H2O ice ih between 0 and 273.16 k." Geochimica Et Cosmochimica Acta **71**(1): 36-45.

Fink, U., et al. (2016). "Investigation into the disparate origin of CO2 and H2O outgassing for comet 67/p." Icarus **277**: 78-97.

Fraser, H. J., et al. (2004). "Using laboratory studies of CO-H2O ices to understand the non-detection of a 2152 cm(-1) (4.647 mu m) band in the spectra of interstellar ices." Monthly Notices of the Royal Astronomical Society **353**(1): 59-68.

Fuchs, G. W., et al. (2006). "Comparative studies of o-2 and n-2 in pure, mixed and layered CO ices." Faraday Discussions **133**: 331-345.

Gerakines, P. A. and Hudson, R. L. (2015). "First infrared band strengths for amorphous CO2, an overlooked component of interstellar ices." Astrophysical Journal Letters **808**(2).

Gerakines, P. A., et al. (1995). "The infrared band strengths of $H_2O$, CO and $CO_2$ in laboratory simulations of astrophysical ice mixtures." Astronomy and Astrophysics **296**(3): 810-818.

Giuliano, B. M., et al. (2014). "Interstellar ice analogs: Band strengths of $H_2O$, $CO_2$, $CH_3OH$, and $NH_3$ in the far-infrared region." Astronomy & Astrophysics **565**.

Greenberg, A. N., et al. (2017). "The effect of $CO_2$ on gases trapping in cometary ices." Monthly Notices of the Royal Astronomical Society **469**: S517-S521.

Guilbert-Lepoutre, A., et al. (2015). "On the evolution of comets." Space Science Reviews **197**(1-4): 271-296.

Guilbert-Lepoutre, A., et al. (2023). "The gateway from centaurs to jupiter-family comets: Thermal and dynamical evolution." Astrophysical Journal **942**(2).

Handa, Y. P., et al. (1986). "High-density amorphous ice. Iii. Thermal properties." The Journal of Chemical Physics **84**(5): 2766-2770.

Heritier, K. L., et al. (2018). "On the origin of molecular oxygen in cometary comae." Nature Communications **9**.

Herny, C., et al. (2021). "New constraints on the chemical composition and outgassing of 67p/churyumov-gerasimenko." Planetary and Space Science **200**: 105194.

Hinkley, S., et al. (2022). "The jwst early release science program for the direct imaging and spectroscopy of exoplanetary systems." Publications of the Astronomical Society of the Pacific **134**(1039).

Hoang, M., et al. (2019). "Two years with comet 67p/churyumov-gerasimenko: $H_2O$, $CO_2$, and CO as seen by the rosina/rtof instrument of rosetta." Astronomy & Astrophysics **630**.

Hoang, M., et al. (2020). "Investigating the rosetta/rtof observations of comet 67p/churyumov-gerasimenko using a comet nucleus model: Influence of dust mantle and trapped CO." A&A **638**: A106.

Hudgins, D. M., et al. (1993). "Midinfrared and far-infrared spectroscopy of ices - optical-constants and integrated absorbances." Astrophysical Journal Supplement Series **86**(2): 713-870.

Isokoski, K., et al. (2014). "Porosity and thermal collapse measurements of $H_2O$, $CH_3OH$, $CO_2$, and $H_2O:CO_2$ ices." Physical Chemistry Chemical Physics **16**(8): 3456-3465.

Jewitt, D. (2009). "The active centaurs." Astronomical Journal **137**(5): 4296-4312.

Jin, M., et al. (2022). "Ice age: Chemodynamical modeling of cha-mms1 to predict new solid-phase species for detection with jwst." Astrophysical Journal **935**(2).

Kouchi, A. and Sirono, S. (2001). "Crystallization heat of impure amorphous h(2)o ice." Geophysical Research Letters **28**(5): 827-830.

Kouchi, A. and Yamamoto, T. (1995). "Cosmoglaciology - evolution of ice in interstellar space and the early solar-system." Progress in Crystal Growth and Characterization of Materials **30**(2-3): 83-108.

Lecar, M., et al. (2006). "On the location of the snow line in a protoplanetary disk." Astrophysical Journal **640**(2): 1115-1118.

Malyk, S., et al. (2007). "Trapping and release of $CO_2$ guest molecules by amorphous ice." Journal of Physical Chemistry A **111**(51): 13365-13370.

Martin-Domenech, R., et al. (2014). "Thermal desorption of circumstellar and cometary ice analogs." Astronomy & Astrophysics **564**.

Mastrapa, R. E., et al. (2013). Amorphous and crystalline H2O-ice. The science of solar system ices. M. S. Gudipati and J. Castillo-Rogez, Springer New York. **356:** 371-408.

Mate, B., et al. (2008). "Ices of CO2/H2O mixtures. Reflection-absorption ir spectroscopy and theoretical calculations." Journal of Physical Chemistry A **112**(3): 457-465.

McClure, M. K., et al. (2015). "Detections of trans-neptunian ice in protoplanetary disks." Astrophysical Journal **799**(2).

Mousis, O., et al. (2021). "Cold traps of hypervolatiles in the protosolar nebula at the origin of the peculiar composition of comet c/2016 r2 (panstarrs)." Planetary Science Journal **2**(2).

Muller, B., et al. (2018). "O-2 signature in thin and thick o-2-H2O ices." Astronomy & Astrophysics **620**.

Muller, B., et al. (2022). "Laboratory spectroscopy of theoretical ices: Predictions for jwst and test for astrochemical models." Astronomy & Astrophysics **668**.

Oberg, K. I., et al. (2011). "The spitzer ice legacy: Ice evolution from cores to protostars." Astrophysical Journal **740**(2): 109.

Oberg, K. I., et al. (2015). "The comet-like composition of a protoplanetary disk as revealed by complex cyanides." Nature **520**(7546): 198-U128.

Owen, J. E. (2014). "Snow lines as probes of turbulent diffusion in protoplanetary disks." Astrophysical Journal Letters **790**(1).

Oza, A. V., et al. (2019). "Dusk over dawn o-2 asymmetry in europa's near-surface atmosphere." Planetary and Space Science **167**: 23-32.

Piso, A. M. A., et al. (2016). "The role of ice compositions for snowlines and the c/n/o ratios in active disks." Astrophysical Journal **833**(2).

Rathbun, J. A., et al. (2010). "Galileo ppr observations of europa: Hotspot detection limits and surface thermal properties." Icarus **210**(2): 763-769.

Rawlings, et al. (2006). "Comparative studies of o-2 and n-2 in pure, mixed and layered CO ices - discussion." Faraday Discussions **133**: 347-374.

Rubin, M., et al. (2020). "On the origin and evolution of the material in 67p/churyumov-gerasimenko." Space Science Reviews **216**(5).

Smith, R. S., et al. (1997a). "Evidence for molecular translational diffusion during the crystallization of amorphous solid water." The Journal of Physical Chemistry B **101**(32): 6123-6126.

Smith, R. S., et al. (1997b). "The molecular volcano: Abrupt ccl 4 desorption driven by the crystallization of amorphous solid water." Phys. Rev. Lett. **79**: 909.

Smith, R. S., et al. (2011). "Crystallization kinetics and excess free energy of H2O and d2o nanoscale films of amorphous solid water." Journal of Physical Chemistry A **115**(23): 5908-5917.

Spencer, J. R., et al. (1995). "Charge-coupled-device spectra of the galilean satellites - molecular-oxygen on ganymede." Journal of Geophysical Research-Planets **100**(E9): 19049-19056.

Taj, S. and McCoustra, M. R. S. (2020). "Thermal desorption of carbon monoxide from model interstellar ice surfaces: Revealing surface heterogeneity." Monthly Notices of the Royal Astronomical Society **498**(2): 1693-1699.

Teolis, B. D., et al. (2009). "Formation, trapping, and ejection of radiolytic o-2 from ion-irradiated water ice studied by sputter depth profiling." <u>Journal of Chemical Physics</u> **130**(13): 9.

Tubiana, C., et al. (2019). "Diurnal variation of dust and gas production in comet 67p/churyumov-gerasimenko at the inbound equinox as seen by osiris and virtis-m on board rosetta." <u>Astronomy & Astrophysics</u> **630**.

van der Marel, N., et al. (2019). "Protoplanetary disk rings and gaps across ages and luminosities." <u>Astrophysical Journal</u> **872**(1).

Walsh, C., et al. (2014). "Complex organic molecules in protoplanetary disks." <u>Astronomy & Astrophysics</u> **563**.

**Figures:**

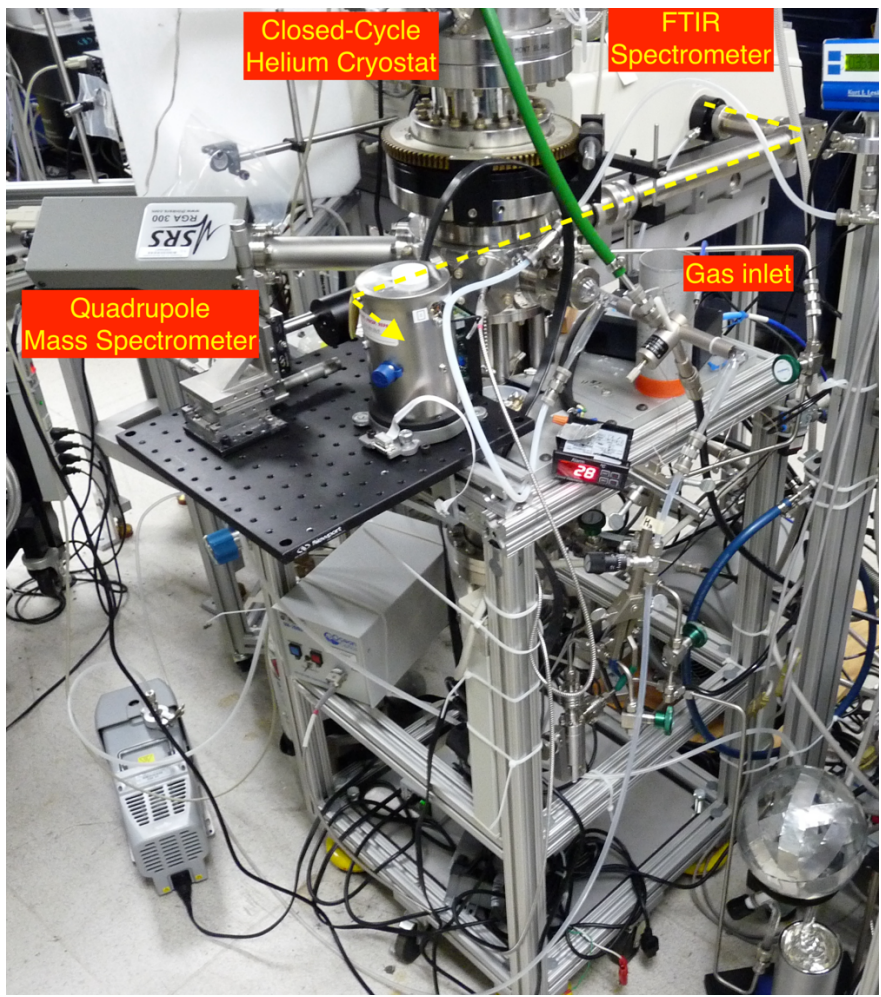

Figure 1: Experimental Station (named Himalaya) that is used to conduct TPD-MS, TPD-IR, and calorimetry experiments at the Ice Spectroscopy Laboratory (ISL) of the Jet Propulsion Laboratory (JPL). FTIR (Thermo-Nicolet 6700) beam is guided through Ni-coated tubes with high-throughput reflecting mirror and passed through the sapphire optical window mounted on a sample holder at the end of the closed-cycle helium cryostat in the vacuum chamber. The transmitted IR beam is focused with a parabolic gold-coated mirror on to the MCT-A detector. Once ice is deposited, the sample is then rotated to be perpendicular to the IR beam and held in that position for the entire TPD experiment. Also mounted on the vacuum chamber is the quadrupole mass spectrometer (Stanford Research Instruments, SRS 300 that is equipped with electron multiplier). In this configuration we have been able to simultaneously record FTIR and mass spectra at a given temperature during the TPD experiments.

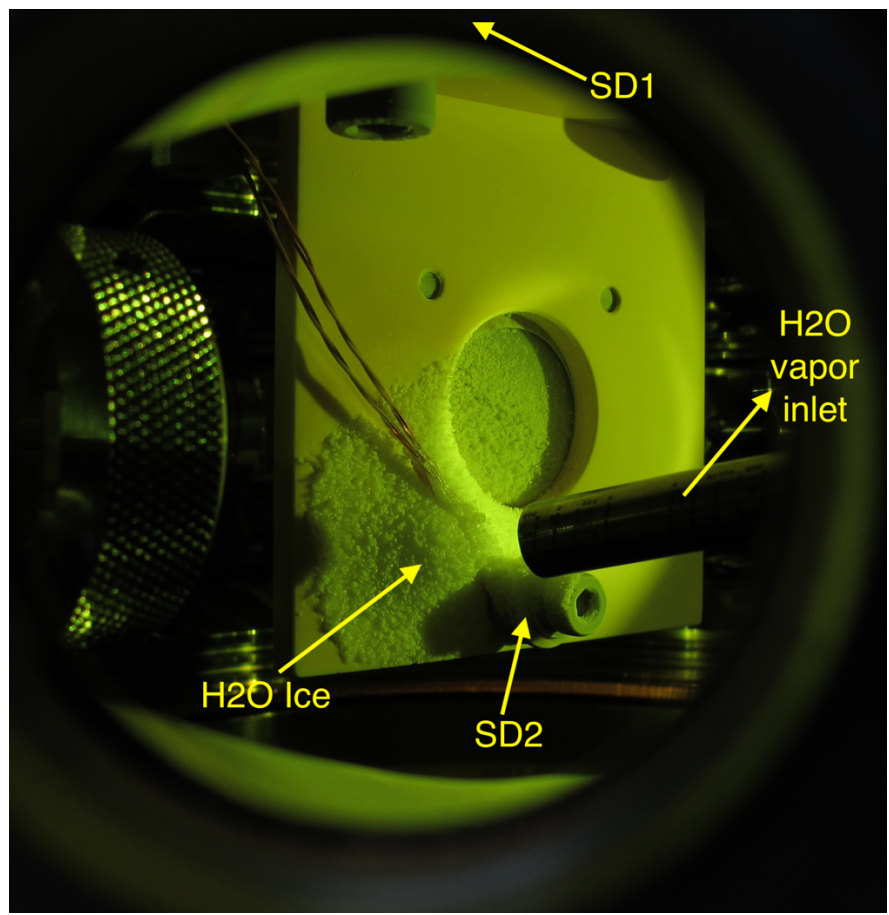

Figure 2: Configuration of Calorimetry on conventional closed-cycle helium cryostat used for ice TPD-MS and TPD-IR experiments. The locations of Silicon Diode (SD) temperature sensors are shown. SD1 that cannot be seen in this photo, is mounted at the top of the ice sample holder that is connected to the end of the closed-cycle helium cryostat. SD2 is mounted at the bottom of the sample holder, which is made of oxygen-free copper. Also seen is the optical window mounted between the copper sample holder, on which normally ice is deposited for TPD experiments. Water vapor inlet is mounted closer to the sample holder and directed to the bottom part to ensure ice deposition is predominantly localized close to SD2.

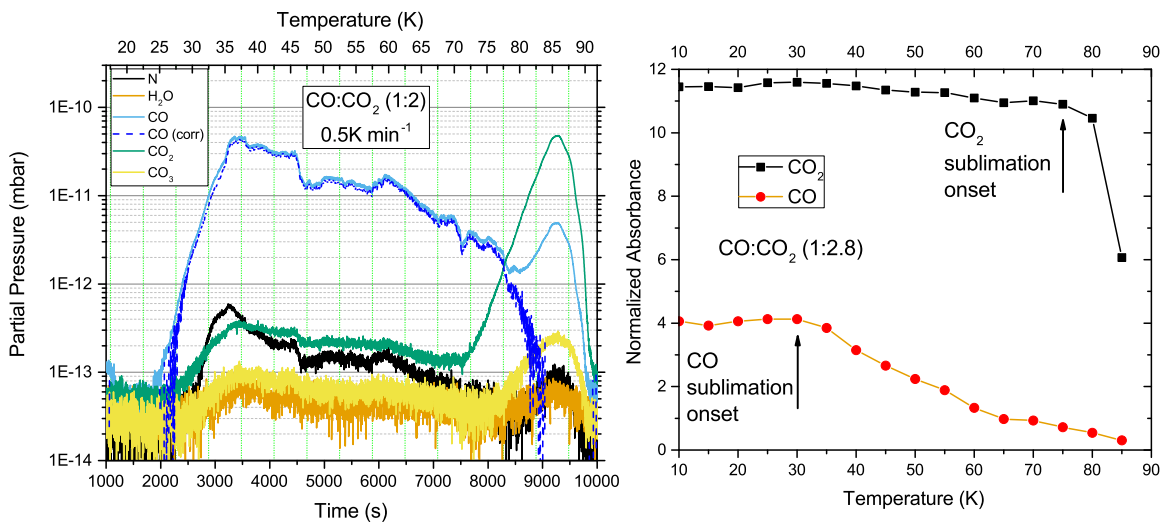

Figure 3: TPD-MS profiles (left) and integrated TPD-IR integrated absorption intensities that are normalized to respective band-strengths (right) of a two-component CO:$CO_2$ ice deposited at 10 K and heated at 0.5 K/min. While the TPD-MS data show an onset of CO sublimation at ~25 K, FTIR spectra show observable sublimation starting after 30 K. $N_2$ sublimation (with N as the proxy) is offset by ~5 K. A small amount of $CO_2$ that is trapped in CO ice is also sublimed along with CO between 30 K and 70 K. Only a small fraction of CO (~10%) remains trapped in $CO_2$ ice at 70 K, the onset of $CO_2$ sublimation, as indicated by the IR data. Please note that $CO_2$ fragmentation in the mass spectrometer results in around 10% of CO, which makes it difficult to quantify the mass spectra, as can be seen in the curve obtained after CO partial pressure correcting for $CO_2$ fragmentation and $N_2$ contribution. We skip this procedure in the rest of the TPD-MS profiles to keep clarity. Here, FTIR spectra are useful for quantification. For mass spectra the initial deposition chamber pressure ratios are given. For IR spectral data, the ratios obtained from integrated and normalized (to their band strengths) relative absorbances are given. Both are from the same ice and are simultaneously measured.

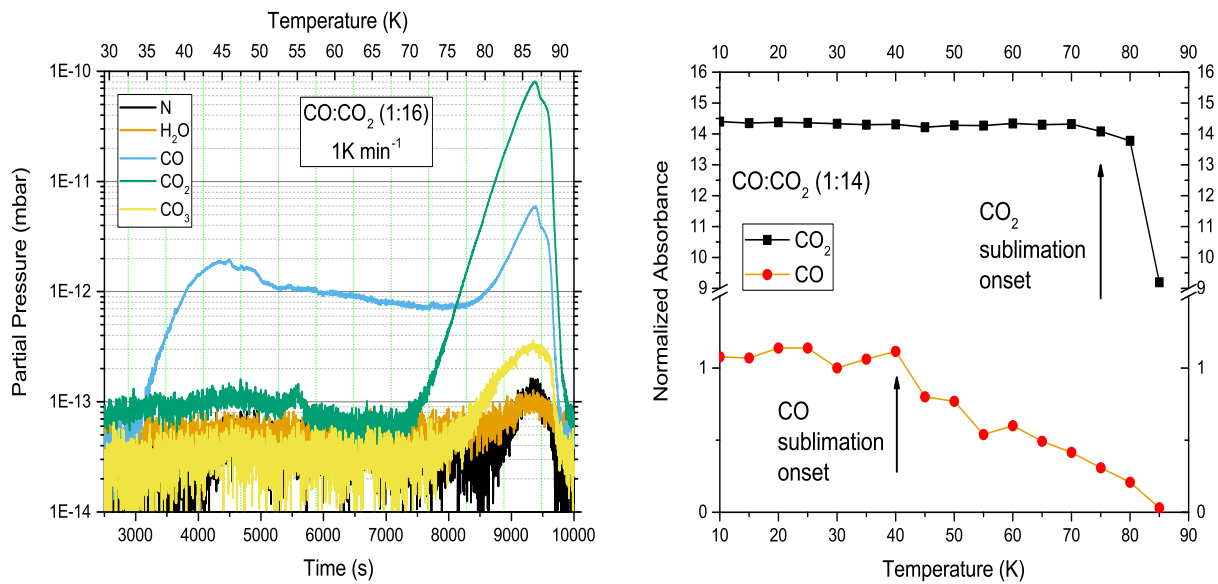

Figure 4: TPD-MS profiles (left) and integrated TPD-IR absorption intensities (right) of two-component CO:$CO_2$ ice deposited at 10 K and heated at 1 K/min. Initial mixing ratio of CO is about 6-7% in $CO_2$ and most of the CO molecules are trapped in $CO_2$ ice. As a result, CO outgassing is delayed by about 10 K from the pure CO ice outgassing (refer to Figure 3). About 3% of CO remained trapped in $CO_2$ ice at 70 K, the onset of $CO_2$ sublimation. For the rest of the information on the figure details, please refer to Figure 3.

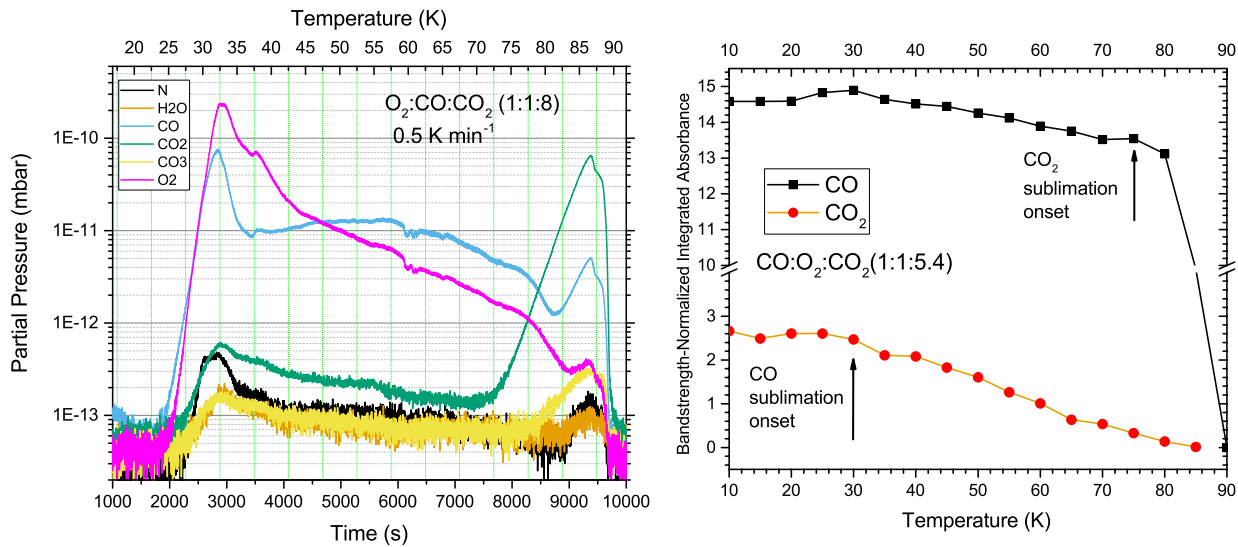

Figure 5: TPD-MS profiles (left) and integrated TPD-IR absorption intensities (right) of three-component CO:$O_2$:$CO_2$ ice deposited at 10 K and heated at 0.5 K/min. In the left figure, it can be seen that the $O_2$ partial pressure is slightly higher between 30 and 35 K and then drops rapidly while CO continues to remain in the vacuum chamber. Though the initial mixing ratio of CO:$CO_2$ is 1:5.4 ($O_2$ is IR inactive and assumed to be about the same mixing ratio as CO), most of the CO sublimes off between 30 K and 70 K, in agreement with the two-component experiments (Figures 3&4), leaving only about 4% CO in $CO_2$ ice at the onset of $CO_2$ sublimation. For the rest of the information on the figure details, please refer to Figure 3.

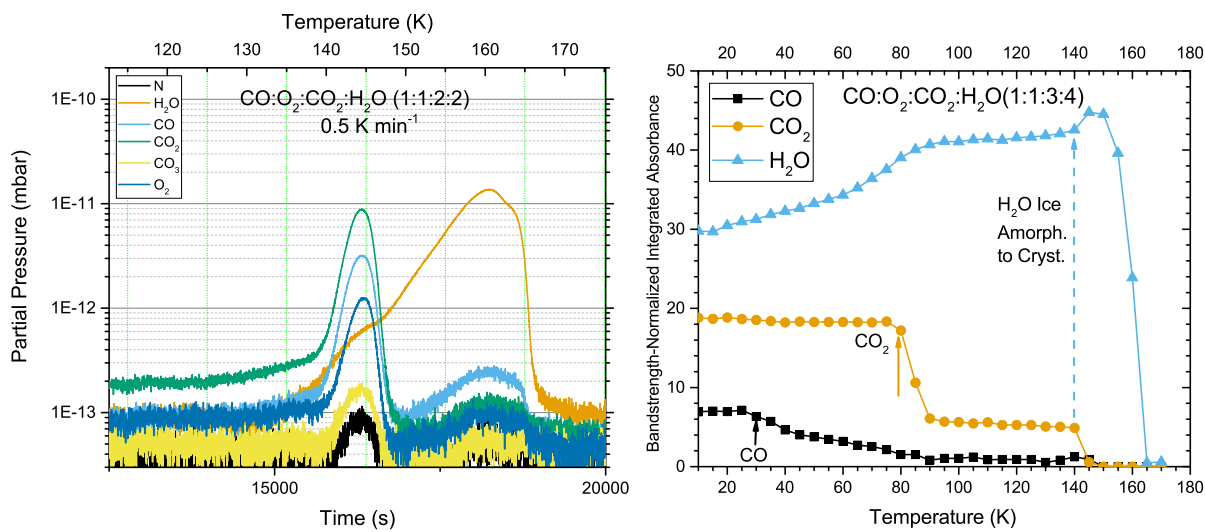

Figure 6: Zoomed-in part of the TPD-MS profiles (left) and integrated TPD-IR absorption intensities (right) of four-component CO:O$_2$:CO$_2$:H$_2$O ice deposited at 10 K and heated at 0.5 K/min. In the left figure, we see between ~137 K and ~147 K a significant outgassing of trapped volatiles in amorphous water ice. In the right figure, we observe similar outgassing onset of CO (hence O$_2$ as well) at 30 K and gradual outgassing until CO$_2$ outgassing that occurs between 70 K and 90 K. Above 90 K, the amount of CO trapped in the amorphous ice is far less than CO$_2$. The data confirm outgassing of almost all CO, O$_2$, and CO$_2$ during crystallization of amorphous ice. During water-ice sublimation only a few percent of these volatiles are still trapped. This experiment demonstrates that if crystalline water ice in astrophysical environments were to be formed, it could accommodate a maximum of 5% of other volatiles in its lattice. If crystalline water ice is observed along with other ices, they must have been formed separately after amorphous water ice has been crystallized (i.e., subjected to temperatures above 120 K for a given amount of time, which is dependent on the maximum exposed temperature below 160 K under vacuum conditions). Please note that water ice integrated absorbance increases constantly and reaches its maximum around 150 K. This is less likely to be due to the increased band strength of water ice with temperature, but rather due to sublimation of a small amount of water from warmer parts of the cryostat and redeposition on the sapphire optical window. For the rest of the information on the figure details, please refer to Figure 3.

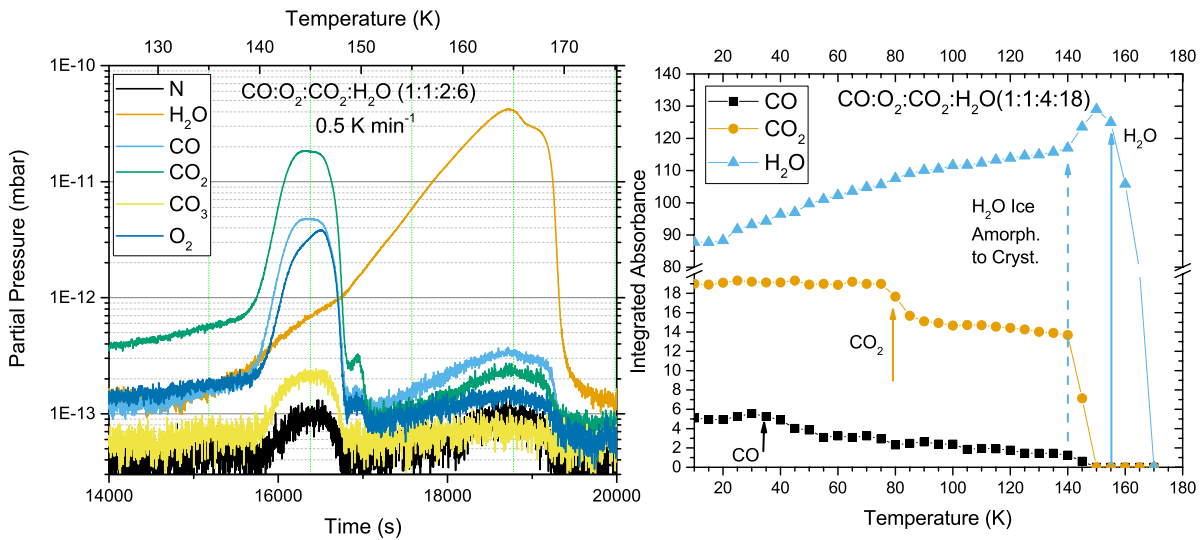

Figure 7: Zoomed-in part of the TPD-MS profiles (left) and integrated TPD-IR absorption intensities (right) of a four-component $CO:O_2:CO_2:H_2O$ ice deposited at 10 K and heated at 0.5 K/min. In this experiment, we increased the mixing ratio of water to test whether there would be any change in the thermal or trapping behavior. We found no changes and as expected, significant amounts of CO ($O_2$) and $CO_2$ were trapped in amorphous ice and released during crystallization. Otherwise, both the two four-component ice types behaved very similarly. Please refer to Figures 6 and 3 for additional details.

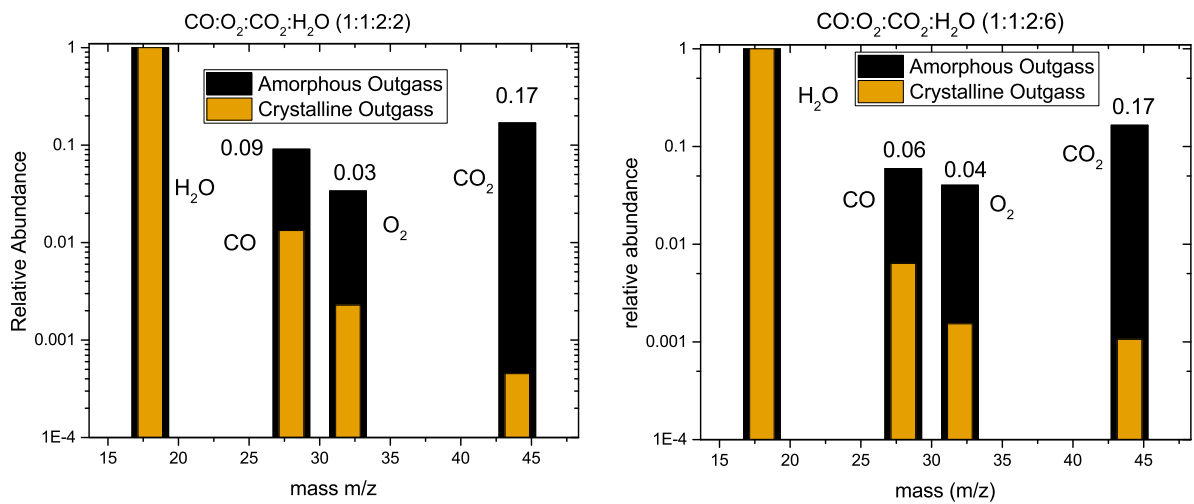

Figure 8: Relative abundances with respect to $H_2O$ calculated from integrated TPD-MS spectra (shown in Figures 4 and 5) of four-component $CO:O_2:CO_2:H_2O$ ice deposited at 10 K and heated at 0.5 K/min. Volatiles CO, $O_2$, and $CO_2$ outgassed during crystallization of amorphous water ice between 130 K and 150 K normalized to the total water content are shown in black bars and the same for crystalline water ice outgassing between 150 K and 190 K are shown in orange bars. In spite of different mixing ratios, in both experiments above 90 K, $CO_2$ is trapped in amorphous ice at an abundance of 17%, while CO and $O_2$ are much lower in abundance (6-9% and 3-4%, respectively). After amorphous ice crystallizes, the total amount of volatiles trapped is reduced to around 1%. Again, our experiments demonstrate that crystalline water ice does not trap significant amounts of volatiles.

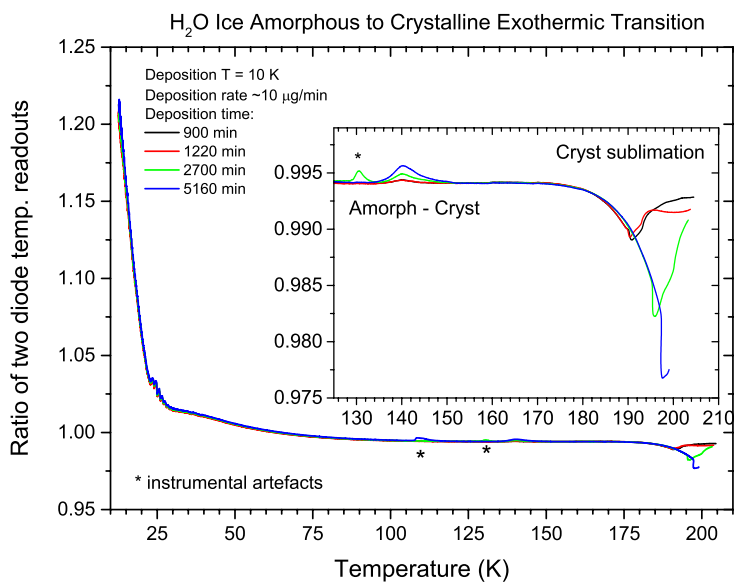

Figure 9: Differential calorimetry during 2K/min pure water ice annealing from amorphous to crystalline to sublimation transitions. Ratio of SD2/SD1 (SD= silicon diode for temperature readout) at various temperatures is plotted for four different ice experiments. SD2 is directly in contact with the ice, whereas SD1 acts as internal reference at the top of the sample holder. For an exothermic phase transition, this ratio is higher and for an endothermic phase transition it is smaller with respect to a common baseline ratio. We consistently see an exothermic transition between 130 and 150 K with maximum at 140 K due to the amorphous to crystalline phase transition. This peak position remains independent of ice thickness, but the transition intensifies with thickness as expected (more ice deposition mass induces a greater heat release and larger differences in the temperatures read at SD1 and SD2). We also observe the endothermic sublimation transition with an onset around 170 K, which also intensifies with thickness. The sublimation onset is the same in all cases, but the sublimation maximum shifts depending on thickness, since it is a surface-driven phenomenon and takes longer for thicker ices to complete the sublimation process. For thinner ice it maximizes at ~ 190 K, and moves to higher temperatures for thicker ice.

# Supplementary Figures

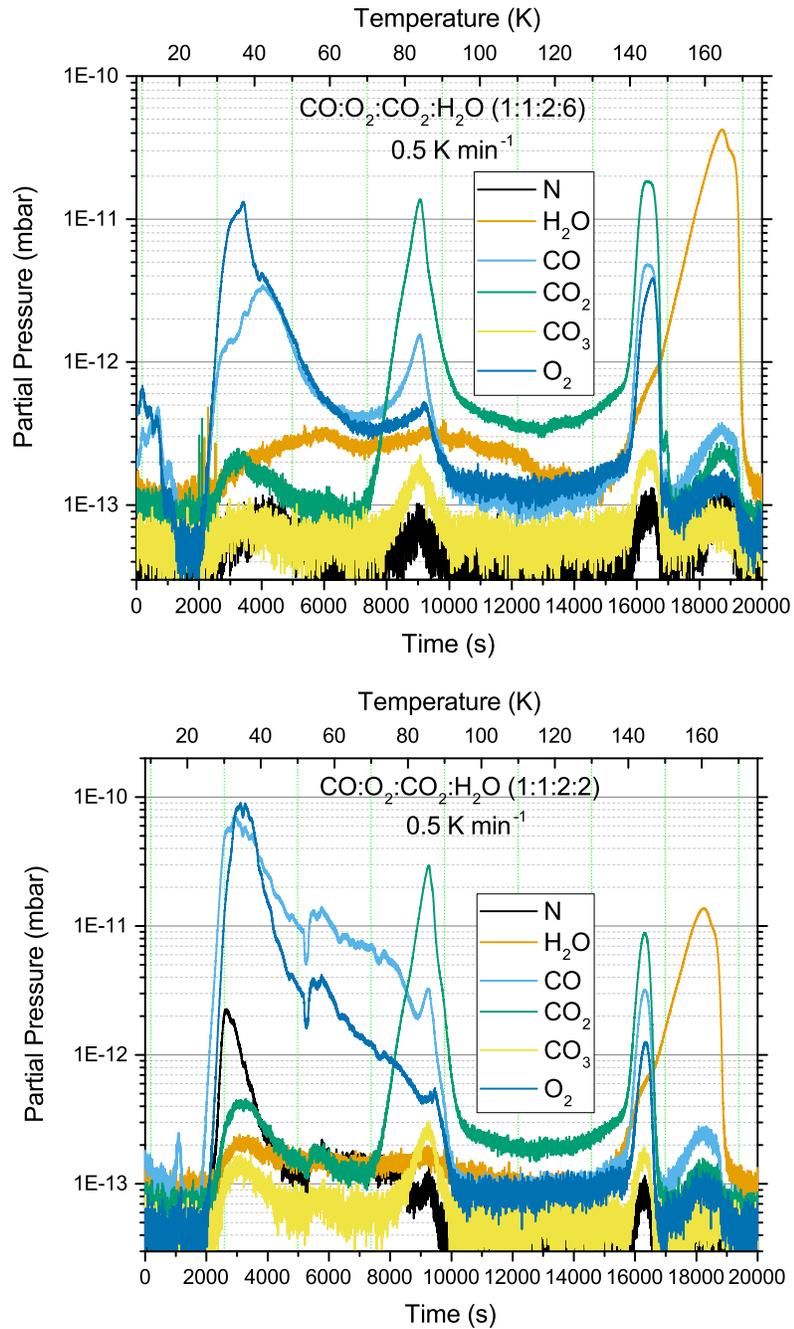

Figure S10: Mass spectral profiles during TPD from 10 K to 190 K of four component mixed molecular ices $CO:O_2:CO_2:H_2O$.

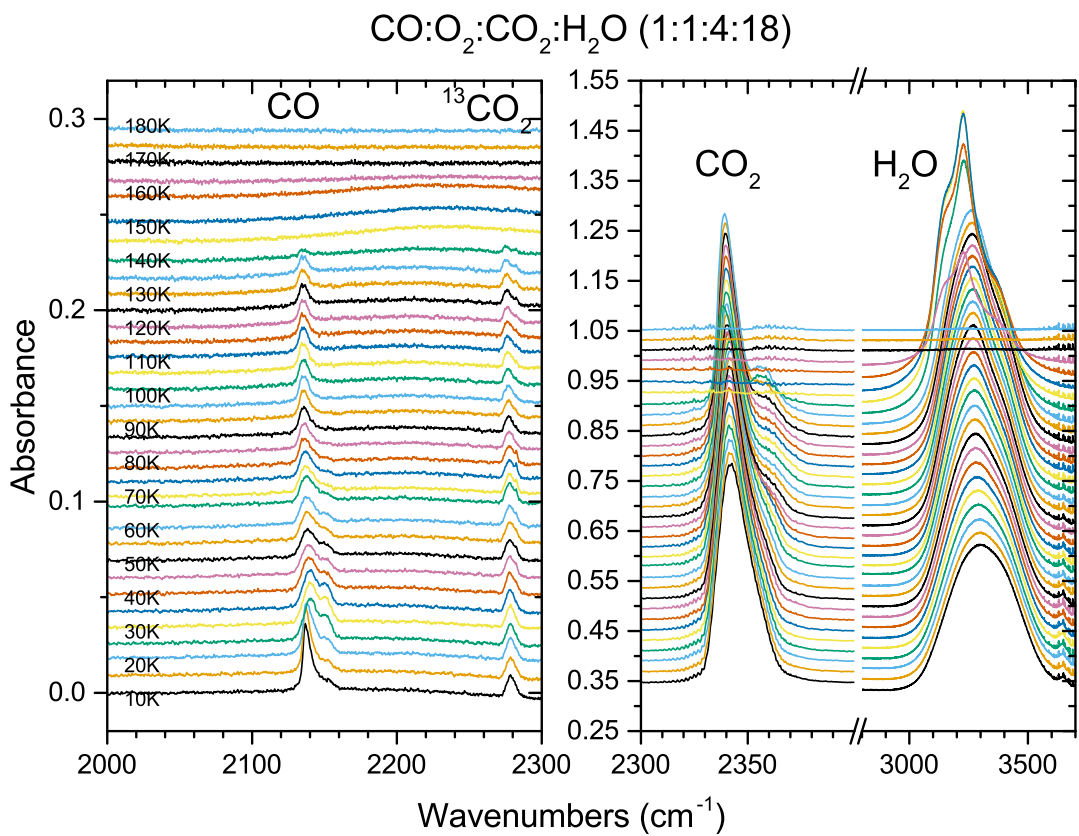

Figure S11: FTIR absorbance spectra of four component CO:$O_2$:$CO_2$:$H_2O$ mixed molecular ice during TPD. Spectra are stacked to provide clarity. CO absorption around 2140 cm$^{-1}$ quickly splits into two components, the higher energy one disappearing by 90 K, indicating the lower energy part should be due to CO trapped in amorphous water-ice. $CO_2$ absorption at 2340 cm$^{-1}$ starts as a single broad band and as the temperature increases a shoulder develops on the higher energy side, which becomes a clear band beyond $CO_2$ ice sublimation. Around 145 K all $CO_2$ sublimes and amorphous water ice band centered around 3250 cm$^{-1}$ develops into a three-band structure, typical of crystalline water ice. Around 185 K, all the water ice sublimes, leaving no residue behind.